\documentclass[twocolumn,prb,aps,amsmath,amssymb,superscriptaddress,showpacs]{revtex4-2}
\usepackage{dcolumn}
\usepackage{ulem}
\usepackage{hyperref}
\usepackage{multirow}
\usepackage{booktabs}
\usepackage{graphicx}
\usepackage{tabularx}
\usepackage{color}
\usepackage{hyperref}
\usepackage{todonotes}
\usepackage{amsmath}
\usepackage{amssymb}
\hypersetup{
    colorlinks=true,
    linkcolor=blue,
    filecolor=magenta,      
    urlcolor=blue,
    citecolor=blue,
    allcolors = blue
}
\makeatletter
\g@addto@macro\normalsize{%
  \setlength\abovedisplayskip{12pt}
  \setlength\belowdisplayskip{12pt}
  \setlength\abovedisplayshortskip{12pt}
  \setlength\belowdisplayshortskip{12pt}
}
\makeatother
\usepackage{physics}
\usepackage{epsfig}
\usepackage{graphics}
\usepackage{float}
\usepackage{xcolor,colortbl}
\usepackage[utf8]{inputenc}
\usepackage[T1]{fontenc}
\usepackage{lmodern} 
\usepackage{soul}
\usepackage[version=4]{mhchem}
\usepackage{soul}

\definecolor{Gray}{rgb}{0.72,0.72,0.98}
\definecolor{LightCyan1}{rgb}{0.83,0.83,0.98}
\definecolor{LightCyan2}{rgb}{0.91,0.92,1}

\begin{document}

\title{Exchange-controlled quantum beats and entanglement \\ 
in an exciton--bimodal-cavity system}

\author{Cesar E. P. Villegas\href{https://orcid.org/0000-0003-2675-1331}{\includegraphics[scale=0.05]{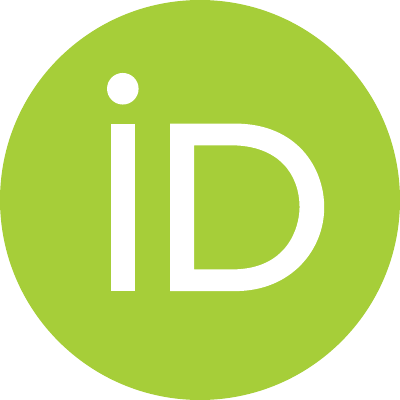}}}
\email[]{cesar.perez@upn.edu.pe}
\affiliation{Physics Department, Federal University of Santa Maria, 97105-900, Santa Maria, RS, Brazil}
\affiliation{Departamento de Ciencias, Universidad Privada del Norte, Lima 15434, Peru.}

\author{Victor P. Brasil\href{https://orcid.org/0009-0009-8203-3976}{\includegraphics[scale=0.05]{orcidid.pdf}}}
\affiliation{Physics Department, 
Federal University of Santa Maria, 97105-900,
Santa Maria, RS, Brazil}

\author{Pedro A. S. Contri\href{https://orcid.org/0009-0007-4519-9745}{\includegraphics[scale=0.05]{orcidid.pdf}}}
\affiliation{Physics Department, 
Federal University of Santa Maria, 97105-900,
Santa Maria, RS, Brazil}

\author{Kevin Lizárraga\href{https://orcid.org/0000-0002-2388-4297}{\includegraphics[scale=0.05]{orcidid.pdf}}}

\affiliation{Center for Quantum Nanoscience, Institute for Basic Science (IBS), Seoul 03760, Republic of Korea}
\affiliation{Ewha Womans University, Seoul 03760, Republic of Korea}

\author{Jonas Maziero\href{https://orcid.org/0000-0002-2872-986X}{\includegraphics[scale=0.05]{orcidid.pdf}}}
\email[]{jonas.maziero@ufsm.br}
\affiliation{Physics Department, 
Federal University of Santa Maria, 97105-900,
Santa Maria, RS, Brazil}

\date{\today}

\begin{abstract}
We theoretically demonstrate tunable population transfer and entanglement generation in a two-exciton system coupled to a bimodal optical cavity, including an exciton-exciton exchange interaction. By using
dressed-state transformations and L\"{o}wdin partitioning, we derive
effective Hamiltonians that identify three dynamical time scales: local
polaritonic oscillations, long-period dressed-state beats, and direct
excitonic exchange. We apply the framework to graphene nanoribbon excitons, using excitonic energies, transition dipoles, and lifetimes obtained from first-principles calculations to parameterize the model. In the small- and moderate-exchange regimes, the dynamics become multiscale, and the beat channel sustains sizable quantum entanglement over long dissipative time windows. In the large-exchange regime, the excitonic Hamiltonian dominates and drives nearly complete population transfer between excitonic states over many coherent cycles. The resulting analytical expressions provide direct design rules connecting the beat period, transfer amplitude, and coherent-cycle number to the exchange interaction, cavity detuning, and light--matter coupling. Our results establish a controllable route for engineering long-lived entanglement and coherent excitonic operations in exciton--bimodal-cavity systems.

\end{abstract}

\maketitle

\section{Introduction}
The design of robust quantum architectures capable of mitigating environmental noise and preserving quantum coherence and quantum entanglement remains one of the most challenging issues in modern quantum optics and quantum information science \cite{awschalom2025challenges,de2021materials}. 
In particular, optical cavities coupled to quantum emitters are attractive platforms for this purpose since they facilitate the tuning of light-matter interactions via the external control of system parameters, allowing the observation of phenomena such as vacuum Rabi oscillations \cite{ridolfo2011all,bose2014all}, polariton formation \cite{luo2023nanophotonics,toffoletti2025coherent}, and long-lived quantum entanglement \cite{wang2021cavity,santos2023generation,kim2025cavity}.

In addition to the standard single-emitter--single-cavity architecture \cite{moehring2007entanglement,kim2018coherent,van2022entangling,fognini2019dephasing,basso2019entanglement,mirhosseini2018superconducting,ferreira2024deterministic}, multi-mode and multi-cavity-QED systems offer additional mechanisms for quantum control, where spatially separated qubits must be coupled in order to transfer quantum states and generate quantum entanglement \cite{skarja1999quantum,ogden2008dynamics,badshah2014dynamics,mckay2015high,amgain2025tripartite}. For instance, inter-cavity coupling can be mediated by photonic tunneling, evanescent-field coupling, optical fibers, or delocalized cavity modes, providing additional degrees of freedom to control the dynamics beyond local light--matter interaction alone \cite{pellizzari1997quantum,cirac1997quantum,serafini2006distributed,skarja1999quantum,badshah2014dynamics}. Inter-cavity coupling can also tune the effective detuning between emitters and delocalized photonic modes, allowing state transfer to occur either via real photon population or through dispersive, virtually populated channels \cite{serafini2006distributed,ogden2008dynamics,zhang2010dynamics}. In coupled-cavity arrays, excitation transfer and multipartite entanglement have been shown to arise from the competition between local emitter--cavity
coupling and intersite photon hopping, with the resulting populations displaying fast and slow oscillations governed by these two energy scales \cite{angelakis2007photon,irish2008polaritonic,badshah2014dynamics}.
These results demonstrate that multimode and multicavity QED systems are promising platforms in which coupling, detuning, and hopping can be used to control coherent transfer and entanglement generation.

Nevertheless, most of these works have focused on atomic ensembles, superconducting qubits, quantum dots, or generic two-level emitters.
Complementary to these platforms, atomically thin semiconductors offer a promising solid-state route because they host strongly bound excitons that couple efficiently to confined electromagnetic fields
\cite{jiang2017scaling,villegas2024screened,schneider2018two,montblanch2021confinement,luo2024strong}, as well as internal spin, valley, or layer degrees of freedom that can be exploited for quantum-state engineering \cite{wilson2021excitons,bao2022light,sun2024twisted}. These properties have motivated increasing interest in exciton-polariton and cavity-exciton systems based on atomically thin materials for quantum technologies~\cite{basov2016polaritons,
helmrich2024cavity,kallioniemi2025van,kipp2025cavity}.
\begin{figure}[ht!]
\centering
\includegraphics[width=1.00\linewidth]{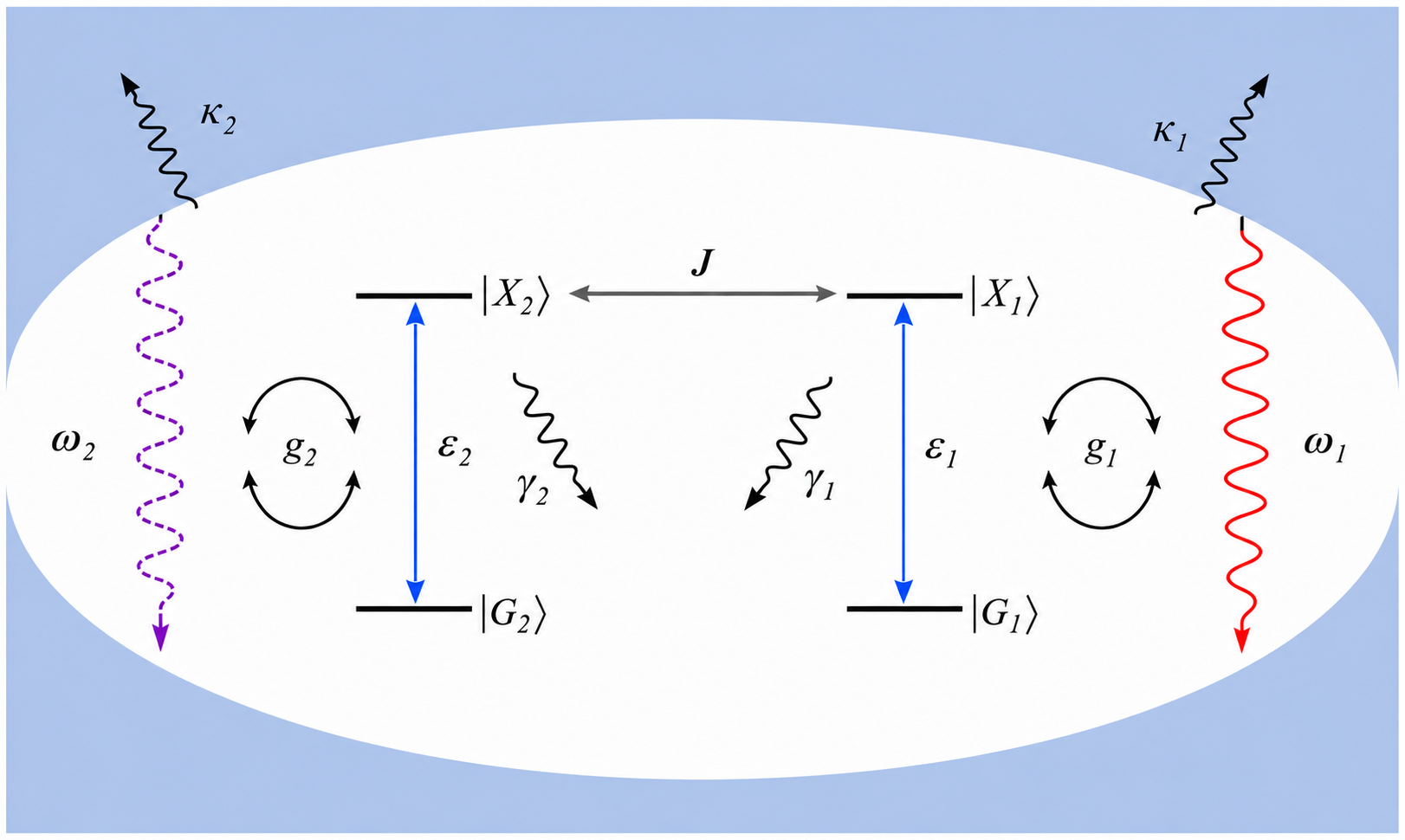}
\caption{Schematic representation of a two-exciton system coupled to a bimodal optical cavity, indicating the exciton ($\varepsilon_{_{1,2}}$), and cavity frequencies ($\omega_{_{1,2}}$), the
exciton--cavity couplings ($g_{1,2}$), as well as the exciton ($\gamma_{1,2}$) and photon ($\kappa_{1,2}$) decay rates.}
\label{fig1}
\end{figure}

Atomically precise one-dimensional nanostructures are especially appealing in this context. For instance, carbon nanotubes have been used to realize exciton-polaritons and excitonic quantum emitters
\cite{graf2017electrical,zhang2026exciton,he2018carbon,chang2026quantum}. Graphene nanoribbons (GNRs) provide another promising platform thanks to recent progress in bottom-up synthesis~\cite{narita2015bottom}, which now
enables the fabrication of edge-precise GNRs with finite and tunable band gaps, strong excitonic effects, and long radiative lifetimes \cite{villegas2024screened,rocha2025tunable}. These features make GNRs attractive building blocks for nanoscale optoelectronics, nanophotonics, and quantum technologies~\cite{lou2024graphene,
gu2022nanographenes,saraswat2021materials,wang2021graphene}.

From a quantum-technology perspective, it is important not only to generate exciton--photon or exciton--exciton quantum correlations, but also to identify regimes where these correlations persist over useful time windows. In this regard, compact analytical descriptions connecting the microscopic Hamiltonian parameters to experimentally relevant observables is essential. Such a description is not only useful for understanding the dynamics, but also provides direct design rules for controlling the duration of coherent oscillations, the amplitude of population transfer, and the robustness of pairwise quantum entanglement.

In this work, we study the coherent population dynamics and entanglement generation in a two-exciton system coupled to a bimodal cavity, including local exciton–cavity coupling and  exciton-exciton exchange interactions. By combining dressed-state transformations with L\"{o}wdin partitioning \cite{lowdin1962studies}, we derive effective Hamiltonians that identify three characteristic dynamical time scales: a local polaritonic period, a long-period dressed beat, and a direct excitonic exchange period. We then develop a unified amplitude formulation that connects these analytical time scales with the bare-state populations and pairwise entanglement generated from a chosen initial state. We apply the framework to graphene nanoribbon excitons, using excitonic energies, transition dipoles, and decay rates obtained from first-principles calculations to parameterize the model.

The results show that small and moderate exchange interactions produce multiscale dynamics in which local photon--exciton oscillations coexist with a long beat channel. In the presence of Lindblad dissipation, this beat channel can sustain extended entanglement concurrence plateaus, indicating that sizable quantum correlations can survive over long dissipative time windows. In the large-exchange regime, the excitonic Hamiltonian becomes dominant and drives nearly complete population transfer between excitonic states over many coherent cycles. These findings establish coupled exciton--bimodal cavity systems as tunable platforms for long-lived entanglement and coherent excitonic operations in nanoscale quantum photonic devices.


\section{Theoretical framework}
\label{sec:framework} 

Here we consider a generic quantum-optical system consisting of two excitons coupled to a bimodal electromagnetic cavity. The exciton states are described by the projectors $|X_1\rangle\langle X_1|$ and $|X_2\rangle\langle X_2|$, with transition energies $\varepsilon_{1}$ and $\varepsilon_{2}$, respectively. The corresponding cavity-mode frequencies are denoted by $\omega_1$ and $\omega_2$.
We assume that each exciton couples predominantly to its corresponding cavity mode, with light-matter coupling strengths $g_1$ and $g_2$. In addition, the two excitons interact through an exchange coupling $J$, arising from Coulomb-mediated spin interactions, which enables coherent excitation transfer between them.
Within the rotating-wave approximation, the Hamiltonian is given by
\begin{align}
 H & = \sum_{\tau
 =1}^{2} \left[ \varepsilon_{\tau} |X_{\tau}\rangle\langle X_{\tau}|  + \omega_{\tau} a_{\tau}^\dagger a_{\tau} + g_{\tau}(\sigma^{+}_{\tau} a_{\tau} + \sigma^{-}_{\tau} a_{\tau}^\dagger) \right] \nonumber \\
 & + J(\hat{\sigma}^{\dagger}_{1}\sigma^{-}_{2}+\sigma^{-}_{1}\sigma^{\dagger}_{2}),
 \label{eq1}
\end{align}
where $a_{\tau}$ and $a^{\dagger}_
{\tau}$ denote the annihilation and creation operators of cavity mode $\tau$, while $\sigma^{-}_{\tau}$ and $\sigma^{\dagger}_{\tau}$ are the lowering and raising operators associated with exciton $\tau$.

To account for the unavoidable losses present in realistic nanocavities, dissipative processes are incorporated using the Lindblad formalism. Specifically, we consider three dissipation channels: exciton recombination with rate $\gamma$, pure exciton dephasing with rate $\gamma_{\phi}$, and photon leakage from the cavity with rate $\kappa$.
The reduced density matrix obeys the master equation \cite{master_eq}:
\begin{equation}
\dot{\rho}=-i[H,\rho]+\sum_{\tau=1}^{2} \left[ \kappa_{\tau} \mathcal{L}_{a_{\tau}}(\rho)+\gamma_\tau\mathcal{L}_{\sigma_\tau^-}(\rho) + \gamma_{\phi_{\tau}}\mathcal{L}_{\sigma_\tau^z}(\rho) \right],
\label{eq2}
\end{equation}
where $\mathcal{L}_O(\rho)=O\rho O^\dagger-\frac{1}{2}\left(O^\dagger O\rho+\rho O^\dagger O\right)$ denotes the standard Lindblad superoperator. Eq.~\eqref{eq2} is solved using the Python-based package \textsc{QuTiP} \cite{lambert2026qutip}, providing direct access to the time evolution of the excitonic and photonic populations, as well as their quantum correlations. The degree of entanglement is quantified by the entanglement concurrence \cite{entanglement}:
\begin{equation}
C(\rho_t)=\max\left[0,\lambda_1-\lambda_2-\lambda_3-\lambda_4 \right],
\label{eq3}
\end{equation}
where $\lambda_i$ are the square roots of the eigenvalues of
\begin{equation}
R_W=\rho_t(\sigma_y\otimes\sigma_y)\rho_t^*(\sigma_y\otimes\sigma_y)
\label{eq4}
\end{equation}
in decreasing order, with $\sigma_y=\begin{bmatrix}0&-i\\i&0\end{bmatrix}$, $\sigma_z=\begin{bmatrix}1&0\\0&-1\end{bmatrix}$ are Pauli matrices, $\sigma^-=\begin{bmatrix}0&1\\0&0\end{bmatrix}$ is the exciton lowering operator and $\rho_t^*$ is the complex conjugate of the density matrix $\rho_t$.

As discussed later, the choice of the excitonic platform is crucial. It should support optically allowed excitations with both large and small oscillator strengths, enabling the competition between coherent coupling and dissipative processes. This interplay ultimately gives rise to a rich variety of quantum phenomena.

Considering the reduced single-excitation subspace,
we adopt the basis $\mathcal{B}=$
$\left\{|C_1\rangle,|C_2\rangle,|X_1\rangle,|X_2\rangle \right\}$,
where $|C_1\rangle$ and $|C_2\rangle$ denote states with a single photon in cavity modes 1 and 2, respectively, and no excitonic excitation, while $|X_1\rangle$ and $|X_2\rangle$ denote states with a single excitonic excitation and no cavity photon. In this basis, the Hamiltonian in Eq.~\eqref{eq1} can be written in the matrix as
\begin{equation}
H
=
\begin{pmatrix}
\omega_1 & 0        & g_1 & 0   \\
0        & \omega_2 & 0   & g_2 \\
g_1      & 0        & \varepsilon_1 & J   \\
0        & g_2      & J   & \varepsilon_2
\end{pmatrix}.
\label{eq5}
\end{equation}
Its eigenvalues can be obtained from the following characteristic equation:
\begin{align}
&0=(\omega_1-\lambda)(\omega_2-\lambda)\left[
(\varepsilon_1-\lambda)(\varepsilon_2-\lambda)-J^2\right]
\nonumber\\
&-g_1^2(\omega_2-\lambda)(\varepsilon_2-\lambda)
-g_2^2(\omega_1-\lambda)(\varepsilon_1-\lambda)+g_1^2g_2^2.
\label{eq6}
\end{align}
Although a closed-form solution exists
for this quartic equation, the resulting expressions involve cumbersome radicals that hide the physical interpretation. Moreover, Eq.~\eqref{eq6} already suggests that the eigenvalues are governed by competing energy scales determined by the relative magnitudes of the exchange interaction $J$, the local exciton-photon couplings $g_1$ and $g_2$, and the detunings between the bare excitonic and photonic states. This observation motivates a treatment in which the problem is decomposed into subsystems corresponding to distinct energy regimes.

\subsection{Excitonic hybridization regime}

Initially, we consider the regime in which the exciton-exciton exchange interaction is the dominant energy scale, i.e., $J \gg g_1, g_2$. In this limit, the cavity modes act as weak perturbations on the relevant dynamical time scale, and the system's dynamics is essentially governed by the excitonic subspace $\{|X_1\rangle,|X_2\rangle\}$, whose Hamiltonian,
\begin{equation}
H_X
=
\begin{pmatrix}
\varepsilon_1 & J \\
J   & \varepsilon_2
\end{pmatrix},
\label{eq7}
\end{equation}
yields the dressed excitonic energies $\varepsilon_{\pm}
=\frac{\varepsilon_1+\varepsilon_2}{2}\pm\frac{\Omega_X}{2}$, where
\begin{equation}
\Omega_X=\sqrt{\Delta_X^2+4J^2},
\label{eq8}
\end{equation}
is the generalized excitonic Rabi frequency, and $\Delta_X = \varepsilon_2-\varepsilon_1$ denotes the excitonic detuning.
The corresponding dressed excitonic eigenstates are given by
\begin{align}
|X_+\rangle&=\sin\theta\,|X_1\rangle+\cos\theta\,|X_2\rangle,
\label{eq9}
\\
|X_-\rangle&=\cos\theta\,|X_1\rangle-\sin\theta\,|X_2\rangle,
\label{eq10}
\end{align}
where the mixing angle satisfies $\tan 2\theta=\frac{2J}{\Delta_X}$.
From the analysis above, the excitonic oscillation period can be obtained analytically as $T_X=
\frac{2\pi\hbar}{\Omega_X}$, which characterizes the coherent population exchange between the two excitons. As will be shown later, it represents the shortest characteristic timescale of the system in which quantum dynamics can occur, offering significant potential for quantum information processing.

\subsection{Local polaritonic regime}
\label{subsec:local_regime_schur}
We now derive the effective Hamiltonian governing the local photon-exciton dynamics within the $\{|C_1\rangle,|X_1\rangle\}$ subspace. This regime is relevant when the system's dynamics occurs near the bare energies of the $C_1$-$X_1$ sector, while the remaining states, $\{|C_2\rangle,|X_2\rangle\}$, contribute primarily through off-resonant virtual processes. Rather than neglecting these external degrees of freedom, we integrate them out using the L\"{o}wdin partitioning procedure \cite{lowdin1962studies}.

To isolate the local dynamics, we reorder the basis as $\mathcal{B}_{P,Q}=$$\left\{|C_1\rangle,|X_1\rangle;|C_2\rangle,|X_2\rangle \right\}$,
where the first two states span the local subspace $P$, while the remaining two span its complementary subspace $Q$. In this reordered basis, the Hamiltonian assumes the block form
\begin{equation}
H
=
\begin{pmatrix}
H_P & V \\
V^{\dagger} & H_Q
\end{pmatrix},
\label{eq11}
\end{equation}
with
\begin{equation}
H_P=
\begin{pmatrix}
\omega_1 & g_1 \\
g_1      & \varepsilon_1
\end{pmatrix},
\,\,
H_Q=
\begin{pmatrix}
\omega_2 & g_2 \\
g_2      & \varepsilon_2
\end{pmatrix},
\,\,
V=
\begin{pmatrix}
0 & 0 \\
0 & J
\end{pmatrix}.
\label{eq12}
\end{equation}
Note that the two subsystems are coupled exclusively through the exchange interaction $J$, which connects the excitonic states $X_1$ and $X_2$.

In this representation, the eigenvalue equation takes the form
\begin{equation}
\begin{pmatrix}
H_P-\lambda I_P & V \\
V^{\dagger} & H_Q-\lambda I_Q
\end{pmatrix}
\begin{pmatrix}
|\psi_P\rangle \\
|\psi_Q\rangle
\end{pmatrix}
=0.
\label{eq13}
\end{equation}
Eq.~\eqref{eq13} yields an effective eigenvalue problem defined solely in the local subspace:
\begin{equation}
H_{\rm eff}^{(L)}(\lambda)|\psi_P\rangle=\lambda|\psi_P\rangle,
\label{eq14}
\end{equation}
where
\begin{equation}
H_{\rm eff}^{(L)}(\lambda)=H_P+V\left(\lambda I_Q-H_Q\right)^{-1}V^{\dagger}.
\label{eq15}
\end{equation}
By performing the matrix algebra and substituting the representations of Eq.~\eqref{eq12} in the effective Hamiltonian above, we obtain
\begin{equation}
H_{\rm eff}^{(L)}(\lambda)
=
\begin{pmatrix}
\omega_1 & g_1 \\
g_1      & \varepsilon_1+\Sigma_L(\lambda)
\end{pmatrix},
\label{eq16}
\end{equation}
where the self-energy is
\begin{equation}
\Sigma_L(\lambda)=J^2\frac{\lambda-\omega_2}{(\lambda-\omega_2)(\lambda-\varepsilon_2)-g_2^2}.
\label{eq17}
\end{equation}
Eq.~\eqref{eq17} clearly shows that the exchange interaction does not merely introduce a constant correction to $\varepsilon_1$. Instead, it generates an energy-dependent self-energy determined by the spectral structure of the remote $C_2$-$X_2$ polaritonic sector.

In the local regime of interest, the relevant eigenvalues satisfy $\lambda \simeq \omega_1 \simeq \varepsilon_1$ and are well separated from the remote $C_2$-$X_2$ sector. Consequently, if the remote exciton-cavity coupling is weak compared with the detuning between the local and remote sectors, the denominator in Eq.~\eqref{eq17} can be approximated as
$(\lambda-\omega_2)(\lambda-\varepsilon_2)$,
yielding the approximate self-energy $\Sigma_L(\lambda)\simeq \frac{J^2}{\lambda-\varepsilon_2}$. 
Evaluating this expression around $\lambda \simeq \varepsilon_1$ gives $\Sigma_L
\simeq -\frac{J^2}{\Delta_X}$,
where $\Delta_X=\varepsilon_2-\varepsilon_1$ is the excitonic detuning.
Therefore, the dominant influence of the remote exciton $X_2$ on the local $C_1$-$X_1$ dynamics is an exchange-induced redshift of the bare exciton energy $\varepsilon_1$. The corresponding static effective Hamiltonian can thus be written as
\begin{equation}
H_L
=
\begin{pmatrix}
\omega_1 & g_1 \\
g_1      & \varepsilon_1-\frac{J^2}{\Delta_X}
\end{pmatrix},
\label{eq18}
\end{equation}
from which one can easily obtain the local polaritonic frequency $\Omega_L=\sqrt{\Delta_L^2+4g_1^2}$, with $\Delta_L=\left(\varepsilon_1-\frac{J^2}{\Delta_X}-\omega_1 \right)$ being the effective local polaritonic detuning.

Several important physical interpretations emerge from this derivation. First, Eq.~\eqref{eq18} shows that the local oscillation frequency is determined not only by the bare exciton--photon detuning, $\varepsilon_1-\omega_1$, but also by virtual exchange processes involving the remote exciton $X_2$. Second, the exchange interaction modifies the local polaritonic dynamics even when the excitation remains predominantly confined to the $C_1$-$X_1$ subsystem. Third, the correction scales as $J^2/\Delta_X$, demonstrating that the local regime is particularly sensitive to the ratio between the excitonic exchange interaction and the excitonic energy separation. Consequently, the frequency $\Omega_L$ encapsulates the combined influence of local photon-exciton coupling and exciton-exciton exchange.

\subsection{Beat-polaritonic regime in the dressed-exciton basis}
\label{subsec:beat_regime}
Now we investigate the regime in which the dynamics is dominated by the pair $\{|C_2\rangle,|X_+\rangle\}$, i.e., the coupling between the second cavity mode and the upper dressed excitonic state. To derive the corresponding effective Hamiltonian, we first express the bare excitonic states in terms of the dressed-state basis. This is readily obtained by inverting Eqs.~\eqref{eq9} and \eqref{eq10}: 
\begin{align}
|X_1\rangle&=\cos\theta\,|X_-\rangle+\sin\theta\,|X_+\rangle,
\label{eq19}
\\
|X_2\rangle&=-\sin\theta\,|X_-\rangle+\cos\theta\,|X_+\rangle.
\label{eq20}
\end{align}

Using the latter transformation, the full Hamiltonian can be rewritten in the following basis
$\mathcal{B}_{\rm d}=\left\{|C_1\rangle,|C_2\rangle,|X_-\rangle,|X_+\rangle \right\}$, yielding
\begin{equation}
H_{\rm d}
=
\begin{pmatrix}
\omega_1 & 0        & g^{-}_{1}  & g^{+}_{1} \\
0        & \omega_2 & -g^{-}_{2} & g^{+}_{2} \\
g^{-}_{1}  & -g^{-}_{2} & \varepsilon_- & 0 \\
g^{+}_{1}  & g^{+}_{2}  & 0   & \varepsilon_+
\end{pmatrix},
\label{eq21}
\end{equation}
where $g^{-}_{1}=g_1\cos\theta$, $g^{+}_{1}=g_1\sin\theta$, $g^{-}_{2}=g_2\sin\theta$, and $g^{+}_{2}=g_2\cos\theta$ represent the dressed coupling terms.  
Eq.~\eqref{eq21} demonstrates that the cavity mode $C_2$ interacts with both dressed excitonic states: the upper branch $X_+$ via the effective coupling $g^{+}_{2}$, and the lower branch $X_-$ through the off-resonant coupling $g^{-}_{2}$. As a result, the near-resonant dynamics are dictated by the dressed excitonic state $X_+$ rather than by the bare exciton $X_2$.

In order to isolate the slow-beat channel, we define the near-resonant subspace $P_B=\left\{|C_2\rangle,|X_+\rangle\right\}$, and its complement
$Q_B=\left\{|C_1\rangle,|X_-\rangle\right\}$. After reordering the dressed basis as
$\{|C_2\rangle,|X_+\rangle;|C_1\rangle,|X_-\rangle\}$, the Hamiltonian takes the following block form:
\begin{equation}
H_{\rm d}
=
\begin{pmatrix}
H_{P_B} & V_B \\
V_B^{\dagger} & H_{Q_B}
\end{pmatrix},
\label{eq22}
\end{equation}
with
\begin{equation}
\begin{aligned}
H_{P_B}
=
\begin{pmatrix}
\omega_2 & g^{+}_{2} \\
g^{+}_{2} & \varepsilon_+
\end{pmatrix},
\,
H_{Q_B}
=
\begin{pmatrix}
\omega_1 & g^{-}_{1} \\
g^{-}_{1} & \varepsilon_-
\end{pmatrix},
\,
V_B
=
\begin{pmatrix}
0 & -g^{-}_{2} \\
g^{+}_{1} & 0
\end{pmatrix}.
\end{aligned}
\label{eq23}
\end{equation}
By following the procedure used in section \ref{subsec:local_regime_schur}, the Hamiltonian acting within the $P_B$ subspace can be derived
\begin{equation}
H_{\rm eff}^{(B)}(\lambda)=H_{P_B}+V_B\left(\lambda I_{Q_B}-H_{Q_B} \right)^{-1}V_B^{\dagger}.
\label{eq24}
\end{equation}
Thus, eliminating the complementary subspace yields an energy-dependent self-energy correction to the effective two-level beat Hamiltonian:
\begin{equation}
H_{\rm eff}^{(B)}(\lambda)
=
H_{P_B}
+
\Sigma_B(\lambda),
\label{eq25}
\end{equation}
with
\begin{equation}
\Sigma_B(\lambda)
=
\frac{1}{D_B(\lambda)}
\begin{pmatrix}
g^{_{-}2}_{2}(\lambda-\omega_1)
&
-g^{_{+}2}_{1} g^{_{+}}_{2}
\\
-g^{_{+}2}_{1} g^{_{+}}_{2}
&
g^{_{+}2}_{1}(\lambda-\varepsilon_-)
\end{pmatrix},
\label{eq26}
\end{equation}
where
$D_B(\lambda)=(\lambda-\omega_1)(\lambda-\varepsilon_-)
-g^{_{-}2}_{1}$.
This self-energy describes virtual transitions from the subspace $P_B=\{|C_2\rangle,|X_+\rangle\}$ to the complementary subspace $Q_B=\{|C_1\rangle,|X_-\rangle\}$ and back. Its contribution to the interaction process is governed by two factors. First, its diagonal part is controlled by the residual couplings connecting the two subspaces, namely $g^{-}_{2}$ and $g^{+}_{1}$ (see Eq.~\eqref{eq23}). For weak or moderate excitonic hybridization, where $\sin\theta\simeq \theta \simeq J/\Delta_{X}$, these residual couplings are suppressed, thereby reducing the coupling to the eliminated sector. Second, the self-energy corrections remain small provided that the residual couplings are much weaker than the energy separation between the two subspaces. More specifically, this requires $g^{-}_{2}, g^{+}_{1} \ll \Delta_{PQ}$, with 
$\Delta_{PQ}\sim\min\left\{|\omega_2-\omega_1|,|\omega_2- 
\varepsilon_-|,|\varepsilon_+-\omega_1|,|\varepsilon_+-\varepsilon_-| \right\}$. Under these conditions, the ${|C_2\rangle,|X_+\rangle}$ channel is spectrally isolated from the complementary subspace, and the self-energy contributes only as a second-order virtual process relative to the direct beat-channel coupling $g^{+}_{2}$. Consequently, to leading order the effective Hamiltonian reduces to
\begin{equation}
H_B
=
\begin{pmatrix}
\omega_2 & g_2\cos\theta \\
g_2\cos\theta      & \varepsilon_+
\end{pmatrix},
\label{eq27}
\end{equation}
which enables the definition of the beat-polaritonic frequency
\begin{equation}
\Omega_B=\sqrt{\Delta_B^2+4g_2^2\cos^2\theta},
\label{eq28}
\end{equation}
where $\cos^2\theta = \frac{1}{2}\left(1+\frac{\Delta_X}{\Omega_X} \right)$ and $\Delta_B=\omega_2-\varepsilon_+$ is the effective beat detuning. The corresponding beat period $T_B=\frac{2\pi\hbar}{\Omega_B}$ becomes particularly long as the dressed-state detuning $\Delta_B$ approaches zero.

The derivation above clarifies the physical origin of the long-period dynamics. It does not originate from the bare coupling between $C_2$ and $X_2$, but rather from the near-resonant hybridization of $C_2$ with the upper dressed excitonic state $X_+$. The exchange interaction $J$ influences this beat frequency by shifting the energy $\varepsilon_+$, and altering the composition of the dressed excitonic state, effectively renormalizing the light--matter coupling via the factor $\cos\theta$. Consequently, the beat period can be engineered either by tuning the exchange interaction or by adjusting the cavity detuning.

\subsection{Unified formulation for populations and entanglement}
\label{subsec:unified_amplitude_formulation}

Now, we introduce a unified amplitude formalism that connects the previously analyzed subspaces with key physical observables, such as state populations and pairwise entanglement concurrence. This framework also explicitly reveals how a given initial state activates the different dynamical channels of the four-state system. We consider the general initial state
\begin{equation}
|\Psi(0)\rangle=a|C_1\rangle+b|X_2\rangle+c|C_2\rangle+d|X_1\rangle,
\label{eq29}
\end{equation}
where $|a|^2+|b|^2+|c|^2+|d|^2=1$. This state simultaneously couples to both the local cavity--exciton subspace,
$\{|C_1\rangle,|X_1\rangle\}$, and the external cavity--exciton subspace,
$\{|C_2\rangle,|X_2\rangle\}$. Consequently, it enables the coexistence of local photon--exciton exchange, excitonic hybridization, and the long-period beat channel.

The analytical approach developed above allows the dynamics to be written as a coherent superposition of the leading-order propagation channels:
\begin{equation}
|\Psi(t)\rangle \simeq U_1(t)\left[a|C_1\rangle+d|X_1\rangle
\right]+U_2(t)\left[c|C_2\rangle+b|X_2\rangle \right],
\label{eq30}
\end{equation}
where $U_1(t)$ is the effective local propagator associated with the
subspace $\{|C_1\rangle,|X_1\rangle\}$, whereas $U_2(t)$ is the effective
external propagator associated with the beat channel involving
$\{|C_2\rangle,|X_+\rangle\}$.

The dynamics of the local subsystem follows directly from the effective Hamiltonian $H_L$ in Eq. (\ref{eq18}). Besides the evolution of the cavity state,
$U_1(t)|C_1\rangle=A(t)|C_1\rangle+ B(t)|X_1\rangle$,
we also obtain the evolution of the initial excitonic state as $U_1(t)|X_1\rangle=
B(t)|C_1\rangle+F(t)|X_1\rangle$. By exponentiating the effective local Hamiltonian, the corresponding amplitudes are given by
\begin{align}
A(t)&=e^{-i\bar{E}_L t/\hbar}\left[\cos\left(\frac{\Omega_L t}{2\hbar}\right)+i\frac{\Delta_L}{\Omega_L}\sin\left(\frac{\Omega_L t}{2\hbar}\right) \right],
\label{eq31}
\\
B(t)&=-ie^{-i\bar{E}_L t/\hbar}\frac{2g_1}{\Omega_L} \sin\left(\frac{\Omega_L t}{2\hbar}\right),
\label{eq32}
\\
F(t)&=e^{-i\bar{E}_L t/\hbar}\left[\cos\left(\frac{\Omega_L t}{2\hbar}\right)
-i\frac{\Delta_L}{\Omega_L}\sin\left(\frac{\Omega_L t}{2\hbar}\right) \right],
\label{eq33}
\end{align}
with $\bar{E}_L=\frac{1}{2}\left(\omega_1+\varepsilon_1-\frac{J^2}{\Delta_X}\right)$.

The external propagation sector is naturally described in the dressed excitonic basis introduced in Eq.~\eqref{eq20}. Since the upper dressed state $|X_+\rangle$ belongs to the beat subspace $\{|C_2\rangle,|X_+\rangle\}$, the beat-channel propagator introduced in Eq.~\eqref{eq30} must satisfy
$U_2(t)|C_2\rangle=E(t)|C_2\rangle
+C(t)|X_+\rangle$ and 
$U_2(t)|X_+\rangle=C(t)|C_2\rangle+
D(t)|X_+\rangle$. The corresponding time-dependent amplitudes are determined by the effective beat Hamiltonian $H_B$ in Eq. \eqref{eq27} and are given by
\begin{align}
C(t)&=-ie^{-i\bar{E}_B t/\hbar}\frac{2g_B}{\Omega_B}
\sin\left(\frac{\Omega_B t}{2\hbar}\right),
\label{eq34}
\\
D(t)&=e^{-i\bar{E}_B t/\hbar}\left[\cos\left(\frac{\Omega_B t}{2\hbar}\right)
+i\frac{\Delta_B}{\Omega_B}\sin\left(\frac{\Omega_B t}{2\hbar}\right) \right],
\label{eq35}
\\
E(t)&=e^{-i\bar{E}_B t/\hbar}\left[\cos\left(\frac{\Omega_B t}{2\hbar}\right)-i\frac{\Delta_B}{\Omega_B}\sin\left(\frac{\Omega_B t}{2\hbar}\right) \right],
\label{eq36}
\end{align}
where $\bar{E}_B=(\omega_2+\varepsilon_+)/2$. 
In addition, the lower dressed exciton evolves according to
$|X_-\rangle\rightarrow
S(t)|X_-\rangle$, where $S(t)=e^{-i\varepsilon_-t/\hbar}$. Substituting this result into Eq.~\eqref{eq20}, the time evolution of the initial bare exciton state $|X_2\rangle$ becomes
\begin{equation}
U_2(t)|X_2\rangle \simeq \cos\theta \left[C(t)|C_2\rangle+D(t)|X_+\rangle \right]-\sin\theta\,S(t)|X_-\rangle .
\label{eq37}
\end{equation}
By substituting the expressions for the time-evolved local and beat states and applying the inverse dressed-state transformation of Eqs.~\eqref{eq9} and \eqref{eq10}, we recover the original basis $\{|C_1\rangle,|C_2\rangle,|X_1\rangle,|X_2\rangle\}$, yielding the following time-dependent state:
\begin{align}
|\Psi(t)\rangle \simeq& \tilde{A}(t)|C_1\rangle+\tilde{C}(t)|C_2\rangle
+\tilde{B}(t)|X_1\rangle+\tilde{D}(t)|X_2\rangle ,
\label{eq38}
\end{align}
where
\begin{align}
\tilde{A}(t)&=aA(t)+dB(t),
\label{eq39}
\\
\tilde{C}(t)&=cE(t)+b\cos\theta\,C(t),
\label{eq40}\\
\tilde{B}(t)&=aB(t)+dF(t)+c\sin\theta\,C(t)+ \nonumber\\
&\quad b\sin\theta\cos\theta \left[D(t)-S(t)\right],
\label{eq41}
\\
\tilde{D}(t)&=c\cos\theta\,C(t)+b\left[ \cos^2\theta\,D(t)
+\sin^2\theta\,S(t)\right].
\label{eq42}
\end{align}

Eq.~\eqref{eq38} constitutes the central analytical result of the generalized unified amplitude formulation when the lower dressed exciton is retained. The terms proportional to $S(t)$ describe the contribution of the initial bare exciton state $|X_2\rangle$ projected onto the lower dressed state $|X_-\rangle$. Although this contribution is off-resonant with the dominant beat channel, it can still influence the bare exciton populations through interference with the upper dressed-state component.

The formulation above is valid when the dynamics can be interpreted in terms of weakly coupled local and beat-polaritonic regimes. This separation, however, breaks down once the exchange interaction becomes the dominant energy scale. Thus, in the moderate and large $J$ regimes, the  excitonic dynamics is governed by the excitonic Hamiltonian $H_X$. Consequently, to make Eq.~\eqref{eq38} remain valid, the excitonic amplitudes, directly obtained from Eq.~\eqref{eq7}, must be replaced by the following expressions
\begin{align}
\tilde{B}(t)&\simeq e^{-i\bar{E}_X t/\hbar}\left[
d\left(\cos\left(\frac{\Omega_X t}{2\hbar}\right)+
i\frac{\Delta_X}{\Omega_X}\sin\left(\frac{\Omega_X t}{2\hbar}\right)
\right) \right. \nonumber \\
&\quad \left.
-i b\frac{2J}{\Omega_X}\sin\left(\frac{\Omega_X t}{2\hbar}\right) \right],
\label{eq43} \\
\tilde{D}(t) &\simeq e^{-i\bar{E}_X t/\hbar}
\left[b\left( \cos\left(\frac{\Omega_X t}{2\hbar}\right)
\right. \right. \nonumber \\
&\quad \left. \left.
-i\frac{\Delta_X}{\Omega_X}\sin\left(\frac{\Omega_X t}{2\hbar}\right)
\right)
-
i d\frac{2J}{\Omega_X}\sin\left(\frac{\Omega_X t}{2\hbar}\right)\right],
\label{eq44}
\end{align}
where $\bar{E}_X=(\varepsilon_1+\varepsilon_2)/2$.
Once the probability amplitudes are determined, the corresponding populations follow directly from $P_{\mu}(t)=|A_{\mu}(t)|^2$,
$A_{\mu}\in\{\tilde{A},\tilde{B},\tilde{C},\tilde{D}\}$. The pairwise entanglement concurrence between the basis states $A_{\mu}$ and $A_{\nu}$ can then be expressed as
\begin{equation}
C_{\mu\nu}(t)=2\left|A_{\mu}(t)A_{\nu}^{*}(t)\right|.
\label{eq45}
\end{equation}
\subsection{Connection to material platforms}
The results concerning the characteristic frequencies and their corresponding dynamical regimes are generally applicable to any platform that exhibits two spectrally addressable excitonic transitions, finite exciton--photon coupling strengths $g_1$ and $g_2$, and an appreciable exchange interaction $J$ between the excitonic states. 
A direct first-principles evaluation of this coupling is nontrivial, because it requires Coulomb matrix elements between correlated excitonic states and depends sensitively on wave-function overlap. Nevertheless, low-dimensional organic semiconductors  and carbon-based nanostructures are expected to enhance Coulomb-mediated exchange effects due to reduced dielectric screening and strong exciton confinement \cite{sugie2023dependence,villegas2024screened}. In this sense, the
singlet-triplet splitting provides a useful order-of-magnitude indicator of the internal exciton exchange energy scale available in such systems, although this splitting should not be interpreted as a specified value of $J$. This estimate is motivated by the fact that microscopic studies of exciton-exciton interactions and exciton landscapes in low-dimensional semiconductors show that exchange contributions can dominate the excitonic Coulomb matrix elements over the relevant momentum range \cite{erkensten2021exciton,qiu2021signatures}. In fact, reported singlet-triplet splittings in conjugated polymeric systems and graphene nanoribbons are on the order of several hundred meV \cite{wilson2000triplet,rocha2025tunable}. Based on these considerations, exchange-like excitonic couplings in the range of tens to
hundreds of meV are physically plausible in strongly confined one-dimensional excitonic platforms. Therefore, in the present work we treat $J$ as a tunable phenomenological parameter and explore the dynamical consequences over this physically motivated range.

\begin{figure}[t!]
\centering
\includegraphics[width=0.90\linewidth]{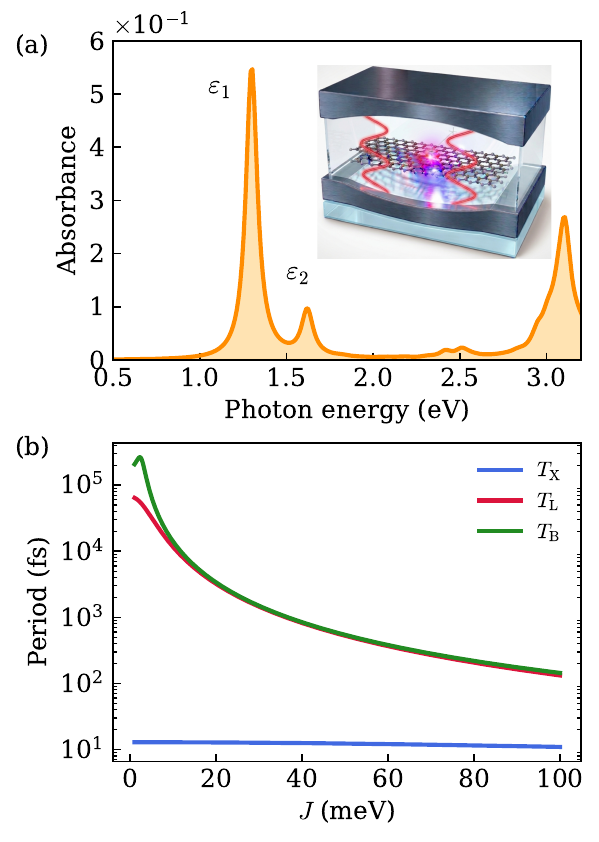}
\caption{(a) Calculated optical absorbance of bilayer 6-AGNR showing the main excitonic states with energies $\varepsilon_1$ and $\varepsilon_2$. The inset shows an schematic representation of a AGNR solid-state platform embedded in a bimodal cavity. (b) Oscillation period for local polaritonic (red line), long-period dressed state beats (green line) and excitonic regime (blue line) as a function of the exciton exchange interaction J.}
\label{fig2}
\end{figure}

Hereafter we consider atomically precise armchair graphene nanoribbons (AGNRs) embedded inside an optical microcavity, as schematically illustrated in the inset of Fig.~\ref{fig2}(a). The cavity is designed to support two confined electromagnetic modes with frequencies $\omega_1$ and $\omega_2$, each capable of coupling to excitonic states. Unlike pristine graphene, AGNRs exhibit well-defined excitonic resonances that can efficiently interact with confined cavity photons.

First-principles calculations have shown that bilayer AGNRs support interlayer excitons, in which the electron and hole are localized in different layers. In contrast to the intralayer excitons found in single-layer AGNRs, these interlayer excitons exhibit distinct transition energies and oscillator strengths. As illustrated in Fig.~\ref{fig2}(a), the large spectral separation between the two lowest-energy excitonic states and the higher excited states allows the excitonic system to be accurately modeled as an effective two-level system. Furthermore, the spatial separation of the electron and hole in interlayer excitons generally leads to longer lifetimes.
The coexistence of intra- and interlayer excitons therefore provides a natural platform for investigating cavity-mediated interactions between multiple excitonic degrees of freedom. Consequently, bilayer AGNRs emerge as a promising platform for exploring the interplay between local polaritonic oscillations, long-period beat dynamics, and direct excitonic exchange interactions.
In order to determine the characteristic time scales, we first calculate the exciton energies, transition dipole moments, and radiative lifetimes by solving the Bethe-Salpeter equation within a first-principles framework. These quantities are obtained from the optical absorption spectrum, which is computed from the imaginary part of the dielectric function:
\begin{equation}
	\epsilon_2(\omega) = \frac{8\pi e^2}{VN_{k}}\sum_S \left| \mu_{_S} \right|^{2}\delta(\omega - \Omega^{S})\,
	\label{eq555}
\end{equation}
where $\Omega^{S}$ is the eigen-energy of the $S$-th exciton. The quantity  $\mu_{_S}=\sum_{vc\mathbf{k}}A^{S}_{vc\mathbf{k}} \textbf{\^{e}} \cdot \left<c\mathbf{k}|\textbf{r}|v\mathbf{k}\right>$ denotes the excitonic transition dipole matrix element, where $\textbf{\^{e}}$ is the light polarization vector, $\textbf{r}$ is the position operator, and $A^{S}_{vc\mathbf{k}}$ is the exciton wavefunction. Here $V$ and $N_{k}$ represent the unit-cell volume and the number of points in the Brillouin zone. Further details of the computational methodology can be found elsewhere \cite{rocha2025tunable}. From the calculated exciton energies and transition dipole moments, the exciton radiative lifetimes and non-radiative lifetimes can be obtained following the procedure described in Refs.~\cite{chen2019ab,villegas2024optical}. Unless otherwise stated, the following parameters are used throughout this work: exciton energies $\varepsilon_{1}=1.30$ eV and $\varepsilon_{2}=1.62$ eV; exciton-cavity coupling $g_{1}=5.32$ GHz and $g_{2}=1.88$ GHz; cavity decay rates 
$\kappa_{1}=0.392$ GHz and $\kappa_{2}=0.489$ GHz; and exciton decay rates $\gamma_{1}=0.208$ GHz and $\gamma_{2}=26.6$ MHz. The pure dephasing rates are assumed to satisfy $\gamma_{\phi_i}=4\gamma_{i}$. The cavity decay rates were estimated from the relation $\kappa_{i}=\omega_{i}Q$, where a cavity quality factor of $Q$=1.25$\times10^{6}$ is assumed.

Fig.~\ref{fig2}(b) compares the characteristic time scales of the system. The excitonic period remains the shortest time scale over the entire range of exchange interactions, giving rise to rapid oscillations in the exciton populations. In contrast, the local and beat periods are substantially longer at small $J$, resulting in well-separated multiscale dynamics. As $J$ increases, both periods decrease and gradually converge, reducing the separation between the local and beat-polaritonic regimes.
As will be shown below, these features are directly reflected in the population dynamics.
\section{Results and discussion}
\begin{figure}[t!]
\centering
\includegraphics[width=0.85\linewidth]{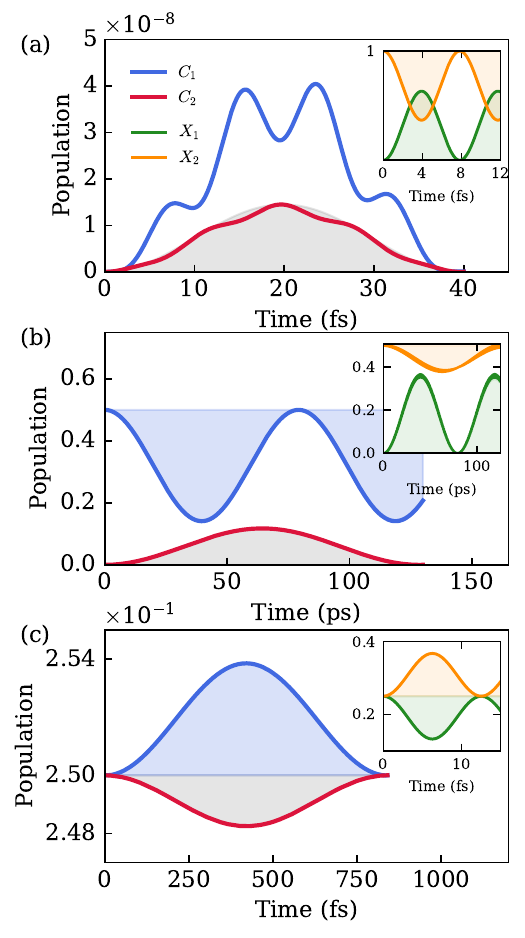}
\caption{Numerical and analytic population dynamics considering three different energy regimes for the exciton exchange. (a) an initial state $|X_2\rangle$ for large $J=0.21$eV. (b) an initial state $(|C_1\rangle+|X_2\rangle)/\sqrt2$ for small $J=3$meV, (c) and initial state $(|C_1\rangle+|X_1\rangle+|C_2\rangle+|X_2\rangle)/2$ for a moderate $J=50$meV. The shaded curves corresponds to the populations obtained with the model hamiltonians while the solid lines corresponds to the numerical solution. All the calculations are carried out assuming the resonant condition. The moderate and strong regime use the excitonic populations in Eq.~\eqref{eq43} and Eq.~\eqref{eq44}. }
\label{fig3}
\end{figure}

In Fig.~\ref{fig3}, we show the effect of the exchange interaction $J$ on
the population dynamics calculated analytically (shaded regions) and numerically (solid lines) for different initial states. The numerical results for the closed system dynamics were also obtained using \textsc{QuTiP} \cite{lambert2026qutip}. The three
representative regimes are well described by the theoretical model, with
only small deviations in the strong-exchange regime shown in
Fig.~\ref{fig3}(a). In this regime, the photon populations exhibit a
dominant long-period oscillation of about $\sim 40$ fs and evolve on the
same global time scale, although their absolute population remains very
small. In contrast, the exciton populations oscillate on a much shorter
time scale, about $\sim 8$ fs, rapidly exchanging population as shown in the
inset. This confirms that, for large $J$, the dynamics is mainly governed by
the excitonic Hamiltonian.

The opposite situation is shown in Figure~\ref{fig3}(b), where the exchange interaction is smaller. In this regime, the photon populations reveal two
different time scales. The faster component, around $\sim 80$ ps, is associated with the local photon--exciton exchange described by Eq.~\eqref{eq18}, whereas the slower modulation, around $\sim 120$ ps, is associated with the beat-channel evolution introduced in Eq.~\eqref{eq27}. The exciton populations, shown in the inset, display
faster oscillations, but their envelope follows the same local-polaritonic and long-period modulation observed in the photon sector.

The intermediate regime using a more general initial state is displayed in Figure~\ref{fig3}(c). There, the populations remain close to their initial values, but small-amplitude oscillations appear as a consequence of the coherent interference between the local and beat propagation channels.
The excellent agreement between the numerical and analytical curves shows that the generalized amplitude formulation captures not only the dominant
population transfer, but also the weak interference corrections produced by the simultaneous activation of different initial components.

\begin{figure}[t!]
\centering
\includegraphics[width=0.89\linewidth]{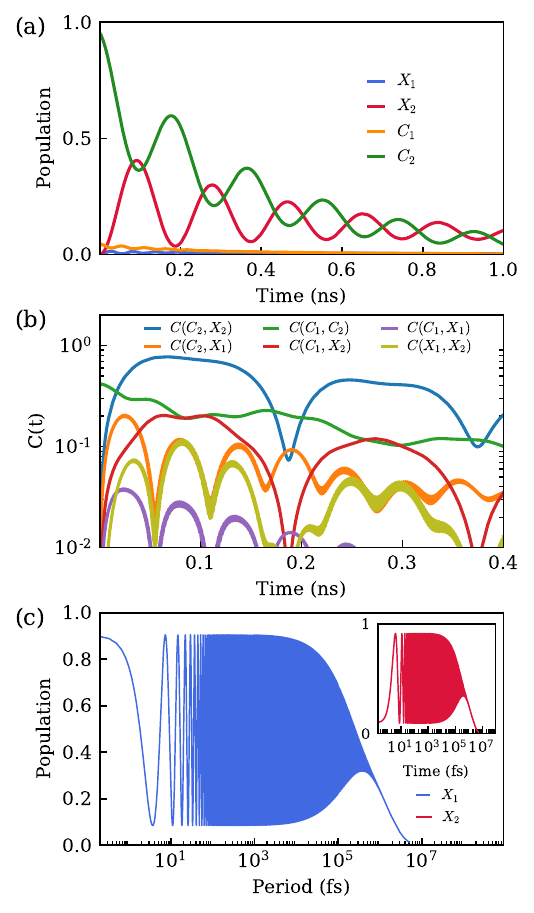}
\caption{(a) Dynanics of photon and exciton populations for the weak $J$ regime corresponding to an initial state $(|C_1\rangle+5|C_2\rangle)/\sqrt26$, and $J=2.35 \mathrm{ meV}$. (b) Corresponding pairwise entanglement concurrence as a function of time.  (c) Populations dynamics for the large $J$ regime considering an initial state $(3|X_1\rangle+|X_2\rangle)/\sqrt10$, and $J=0.23 \mathrm{ eV}$. In all cases the cavity-exciton detunings are fixed to $\omega_1-\varepsilon_1=2g_1$, and $\omega_2-\varepsilon_2=g_2/5$. }
\label{fig4}
\end{figure}
\begin{figure*}[ht!]
\centering
\includegraphics[width=1.0\linewidth]{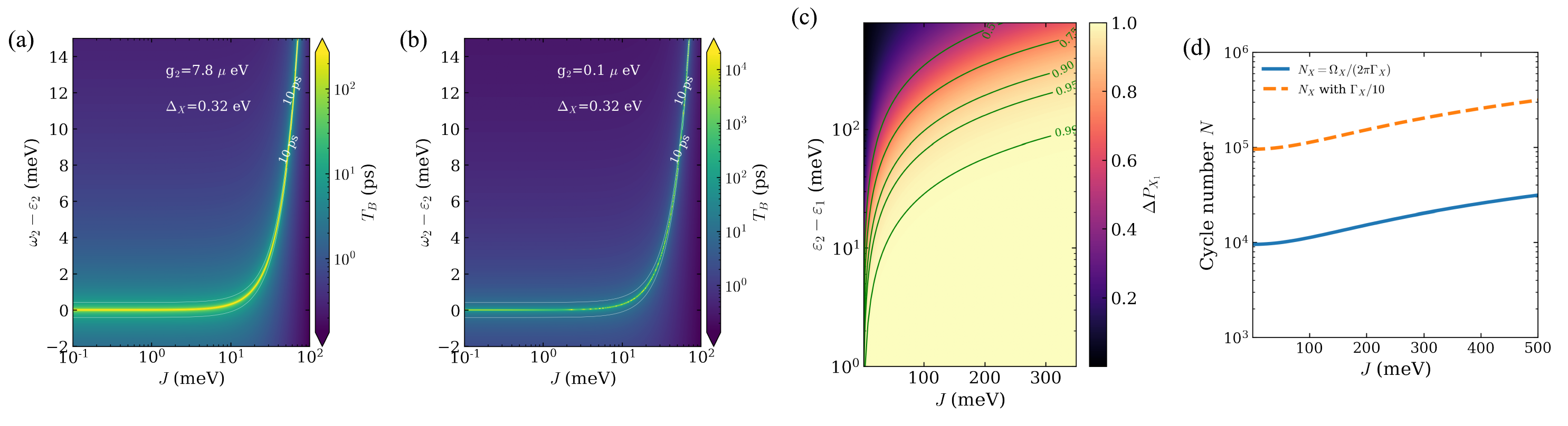}
\caption{Engineering maps for the characteristic time scales and transfer efficiency of the two-exciton--bimodal-cavity system. (a,b) Long beat period $T_B$ as a function of the exciton exchange
interaction $J$ and the external cavity detuning $\omega_2-\varepsilon_2$
for two values of the external exciton--cavity coupling: (a) $g_2=7.8~\mu{\rm eV}$ and (b) $g_2=0.1~\mu{\rm eV}$. (c) Maximum excitonic population swing, $\Delta P_{X_1}=P_{X_1}^{\max}-P_{X_1}^{\min}$, as a function of $J$ and the excitonic detuning $\Delta_X=\varepsilon_2-\varepsilon_1$, assuming an optimized initial state.
(d) Excitonic coherent-cycle number $N=T_2/T_X$ as a function of $J$ under resonant exciton--cavity conditions. The solid curve corresponds to the estimated coherence rate, while the dashed curve illustrates a tenfold
improvement in the excitonic coherence time.
}
\label{fig5}
\end{figure*}

Next, we study the population dynamics in the presence of dissipation and dephasing, described using the Lindblad master equation. For small $J$, Fig.~\ref{fig4}(a) shows that the population, initially predominant in the cavity mode $C_2$, is gradually transferred to the other degrees of freedom while being damped by the environment. The oscillatory pattern combines the effects of the different dynamical channels discussed in the closed-system analysis. Although dissipation progressively reduces the total excitation probability, population exchange remains visible over several hundred picoseconds, indicating that the coherent dynamics are not immediately washed out by photon leakage, exciton decay, and exciton dephasing.

The corresponding pairwise entanglement concurrences are shown in Fig.~\ref{fig4}(b). Remarkably, an extended plateau is observed in the concurrence involving the cavity mode $C_2$ and the excitonic state $X_2$. 
It is worthwhile observing that what we mean, above and hereafter, by entanglement plateau is a time interval with entanglement approximately constant for the unitary dynamics, and that decays slowly for the open system dynamics.
This plateau indicates that, even in the presence of losses, the system can preserve sizable quantum correlations over a time window much longer than the fastest oscillation
periods. The origin of this behavior can be understood from the generalized beat-polaritonic frequency in Eq.~\eqref{eq28}. For the particular condition $\Delta_B=2g_2\cos\theta$, one obtains
$\Omega_B=2\sqrt{2}g_2\cos\theta$. In this regime, the effective beat detuning and the $C_2$--$X_+$ interaction compete on comparable energy scales, suppressing complete energy exchange and generating a slowly varying
interference between the relevant pairwise states. The plateau can therefore be interpreted as a detuning-assisted coherence-preserving mechanism tied to the beat-polaritonic channel.

Furthermore, the smaller entanglement concurrences exhibit faster modulations and stronger decay, reflecting their stronger dependence on rapidly oscillating or weakly populated components. Therefore, the open-system dynamics reveal a hierarchy
of correlations: some pairwise channels decay rapidly, while others remain more robust because they are connected to the slow beat dynamics under the condition $\Delta_B=2g_2\cos\theta$.

To study population transfer under dissipation, Fig.~\ref{fig4}(c) shows the large-$J$ regime, where the dynamics is dominated by the excitonic
Hamiltonian. In this case, the population undergoes nearly complete oscillations between the two excitonic states, approaching the ideal transfer range from nearly one to nearly zero. These oscillations persist for many cycles before dissipation substantially suppresses the signal. This behavior is particularly relevant for coherent-control applications. Although the
exchange-dominated regime does not by itself define a specific quantum gate, it provides a necessary ingredient for gate-like coherent operations, since controlled population transfer must be completed before losses and dephasing
destroy the phase coherence. Thus, the exchange interaction controls not only the characteristic periods of the closed system, but also the robustness and usefulness of the open-system dynamics.

Given the intertwined dependence of the two-exciton--bimodal-cavity
parameters on both the dynamical time scales and the efficiency of
population transfer, these quantities can be engineered through suitable
choices of coupling, detuning, and exchange interaction. In
Fig.~\ref{fig5}(a)--(b), we show the beat period as a function of the
exchange interaction $J$ and the cavity detuning $\omega_2-\varepsilon_2$
for two different values of the external exciton--cavity coupling $g_2$.
We see that reducing $g_2$ substantially increases the long
beat period, by up to two orders of magnitude. This occurs because weaker
external coupling slows down the exchange between the cavity mode and the
upper dressed exciton, producing longer-lived beat oscillations. The bright
resonance line identifies the region where the beat period is maximized,
showing that both cavity detuning and exchange interaction can be used as
control knobs.

From the perspective of practical applications, this tunability is relevant
because it provides direct control over the time window in which coherent
population and entanglement dynamics can occur. In particular, small $g_2$
and proper detuning favor slowly varying beat dynamics, while larger
couplings can be used when faster coherent operations are desired.

Figure~\ref{fig5}(c) shows the maximum achievable excitonic population
swing as a function of $J$ and the excitonic detuning
$\Delta_X=\varepsilon_2-\varepsilon_1$, assuming that the initial state is
optimized. The population-transfer amplitude approaches unity over a broad
region of parameter space. Therefore, the system can be engineered so that
the excitonic population oscillates almost completely between the two
excitonic states. This provides a direct route toward controlled population
inversion: by increasing $J$ or reducing $\Delta_X$, the dynamics approach
the ideal limit in which the population can be switched from nearly zero to
nearly one.

Finally, in Fig.~\ref{fig5}(d), we estimate the excitonic coherent-cycle
number,
$N=\frac{T_2}{T_X}=\frac{\Omega_X}{2\pi\Gamma_X^{\rm coh}}$,
where the effective coherence rate is approximated as
$\Gamma_X^{\rm coh}\simeq \frac{\gamma_1+\gamma_2}{2}+\gamma_{1\phi}+\gamma_{2\phi}$.
This quantity measures how many coherent excitonic oscillations can occur
before dissipation and dephasing dominate. The large values of $N$ indicate
that the exchange-driven channel can survive over many coherent cycles,
which is essential for coherent-control applications. The dashed curve
illustrates that improving the excitonic coherence rates further increases
the number of accessible cycles, emphasizing the importance of material
quality, radiative lifetime, and dephasing suppression.

Overall, Fig.~\ref{fig5} shows that the proposed system offers three
complementary levels of control. The beat period can be tuned through the
external light--matter coupling and cavity detuning, the excitonic transfer
amplitude can be optimized through the exchange interaction and excitonic
detuning, and the coherent-cycle number can be enhanced by improving the
exciton coherence time. Together, these results indicate that
two-exciton--bimodal-cavity systems can support long-lived beat dynamics,
nearly complete excitonic population transfer, and parameter-engineered
coherent time scales. These features make them promising platforms for
controllable excitonic and polaritonic operations in nanoscale quantum
photonic devices.

\section{Conclusions}
In summary, we have investigated population transfer and quantum entanglement generation in a two-exciton system coupled to a bimodal optical cavity. By combining dressed-state transformations with L\"{o}wdin partitioning, we derived effective Hamiltonians that isolate the main dynamical channels of the system. This approach allowed us to identify three characteristic time scales: a local polaritonic period, a long-period dressed-state beat, and a direct excitonic exchange period. These time scales provide a clear physical interpretation of the multiscale dynamics observed in the full numerical solution.

We showed that the exchange interaction $J$ plays a central role in controlling the dynamics. For small exchange interactions, the system exhibits a coexistence of local photon-exciton oscillations and a slow beat channel involving the cavity $|C_2\rangle$  and the upper dressed exciton. The unified amplitude formulation reproduces the numerical population dynamics and shows how different initial states activate local, beat-polaritonic, and excitonic contributions. In addition, in the presence of dissipation, the beat channel can sustain extended entanglement concurrence plateaus, indicating that sizable pairwise quantum correlations
can survive over long dissipative time windows.
Interestingly, in the large-exchange regime, the dynamics becomes dominated by the
excitonic Hamiltonian. In this case, the system supports nearly complete
population transfer between the two excitonic states over many coherent
cycles before dissipation suppresses the signal. This behavior highlights
the usefulness of the exchange interaction as a control parameter for coherent excitonic operations.
Applied to graphene nanoribbon excitons parameterized from first-principles calculations, these results establish exciton--bimodal-cavity systems as promising nanoscale platforms for long-lived quantum entanglement, coherent population transfer, and parameter-engineered quantum dynamics.
 
\section{Acknowledgements}
This work was supported by the National Council for Scientific and Technological Development (CNPq), Grants No. 300083/2025-4 and No. 132266/2025-3, by the National Institute of Science and Technology in Applied Quantum Computing, Grants No. 408884/2024-0 and No. 132191/2026-1 and by the Research Support Foundation for the state of Rio Grande do Sul, Grant. No. 25/2551-0002608-3. K.L. acknowledges the support from the Institute for Basic Science under grant IBS-R027-
D1.

\bibliography{paper}

\end{document}